\documentclass[10pt,journal,onecolumn]{IEEEtran}
\IEEEoverridecommandlockouts
\usepackage{cite}
\usepackage{amsmath,amssymb,amsfonts}
\usepackage{textcomp}
\usepackage{xcolor}
\def\BibTeX{{\rm B\kern-.05em{\sc i\kern-.025em b}\kern-.08em
    T\kern-.1667em\lower.7ex\hbox{E}\kern-.125emX}}

\usepackage{algpseudocode}
\usepackage{algorithm}
\mathchardef\mhyphen="2D % Define a "math hyphen"
\usepackage{empheq}
\usepackage{graphicx}
\usepackage{tabularx}
\usepackage{multirow}
\usepackage{mathtools} 
\usepackage{float}
\usepackage{amsmath,amssymb,lipsum}
\usepackage{color}
\usepackage{colortbl}
\usepackage{tikz}
\usepackage{float}
\usepackage{amsmath}
\usepackage{multirow}
\usepackage{algorithm}
\usepackage{algpseudocode}
\usepackage{subfig}
\usepackage[bookmarks=false]{hyperref}

\begin{document}

\title{Vulnerability Coverage for Secure Configuration\\
\iffalse
{\footnotesize \textsuperscript{*}Note: Sub-titles are not captured in Xplore and
should not be used}
\thanks{Identify applicable funding agency here. If none, delete this.}
\fi
}

\author{\IEEEauthorblockN{Shuvalaxmi Dass and Akbar Siami Namin} \\
\IEEEauthorblockA{\textit{Computer Science Department} \\
\textit{Texas Tech University}\\
%Lubbock, USA \\
shuva93.dass, akbar.namin@ttu.edu}
}

\maketitle

\begin{abstract}
%The conventional software applications and tools are configurable artifacts. These applications are often equipped with a vast number of parameters that can take in different values. The security and resilience of these applications against some known vulnerabilities rely heavily on how certain parameters are tweaked with their different possible settings. Hence, an instance of  configuration settings may lead to either strengthening or weakening the security. From software testing point of view, it is impractical to conduct comprehensive and exhaustive testing of all possible enumerations of the parameters of these tools. Furthermore, due to the presence of enormous number of vulnerabilities reported, it is also impractical to test the underlying software against all vulnerability. The research question we like to address is whether the software under test is robust and secure against these vulnerabilities. 
We present a novel idea on adequacy testing called  ``{vulnerability coverage}.'' The introduced coverage measure  examines the underlying software for the presence of certain classes of vulnerabilities often found in the National Vulnerability Database (NVD) website. The thoroughness of the test input generation procedure is performed through the adaptation of  evolutionary algorithms namely Genetic Algorithms (GA) and Particle Swarm Optimization (PSO). The methodology utilizes the Common Vulnerability Scoring System (CVSS), a free and open industry standard for assessing the severity of computer system security vulnerabilities, as a fitness measure for test inputs generation.  The outcomes of these evolutionary algorithms are then evaluated in order to identify the vulnerabilities that match a class of vulnerability patterns for testing purposes. %In this paper, we compare and report the  performance results of two evolutionary algorithms, namely Genetic Algorithms (GA) and Particle Swarm Optimization (PSO), in generating the vulnerability pattern vectors.     
\iffalse
Mainstream software applications and tools are configurable platforms with an enormous number of parameters along with their values. Certain settings and possible interactions between these parameters may harden (or soften) the security and robustness of these applications against some known vulnerabilities. However, the large number of vulnerabilities reported and associated with these tools make exhaustive testing of these tools infeasible against these vulnerabilities. As an instance of general software testing problem, the research question to address is whether the software under test is robust and secure against these vulnerabilities. This paper introduces the idea of ``{\it vulnerability coverage}'', a concept to adequately test a given application for a certain classes of vulnerabilities, as reported by National Vulnerability Database (NVD). The deriving idea is to utilize the Common Vulnerability Scoring System (CVSS), as a means to measure the fitness of test inputs generated by evolutionary algorithms, and then through pattern matching, identify vulnerabilities that match the generated vulnerability vectors and test the program under test for those identified vulnerabilities. We report the performance of two evolutionary algorithms, namely Genetic Algorithms (GA) and Particle Swarm Optimization (PSO), in generating the vulnerability pattern vectors.  
\fi
\end{abstract}


\section{Introduction}
\label{sec:introduction}

%The enterprise software applications are often shipped with a large number of options and features. As a result, the clients of such software systems are able to customize the configuration of their software systems with respect to their needs. The configuration of the settings can lead to variations in the functionalities that they are responsible for. As a motivating example, MySQL features over 600 setting parameters that can be utilized to configure the database. These parameters are  categorized into three groups namely: Server Option, System Variable, and Status Variable Reference \cite{Mysql}. These parameters when configured properly offer great functionalities to their users; whereas, an improper configuration either caused by a human error or an attacker can create flaws in the system making it prone to vulnerabilities that could result in great deal of loss and misuse of information.

The National Vulnerability Database (NVD) \cite{NVD} lists over $1,644$ instances of  vulnerabilities identified by their unique CVE (Common Vulnerabilities and Exposures) numbers. Some of these vulnerabilities exist partly due to improper settings of configuration parameters that govern the functionality of the given software. Testing software applications against these vulnerabilities can become tedious, and as a result, infeasible if the test has to target all the possible settings of a configuration in order to check the vulnerability of the software against some known attacks. This calls for a systematic configuration testing framework and mechanism in order to mitigate the efforts put into inspecting the given software by identifying a narrowed down set of configuration test inputs. However, from vulnerability perspective, it is not easy to check whether the tests generated for testing examines the software system against certain classes of vulnerabilities. Hence, it makes it impractical for the administrator to exercise the given software under test against any vulnerability reported in NVD. %Hence, the research question that needs to be addressed is whether it is possible for the administrator to identify a subclass of vulnerabilities that represent classes of vulnerabilities and thus focus on selected cases.

To illustrate the overall mechanism of the proposed adequacy coverage, consider an example for vulnerabilities in MySQL. The ``high'' severity level of the MySQL CVE-2019-12463 vulnerability is $8.8$ out of 10. There are several other vulnerabilities reported for MySQL with similar patterns as vulnerability vector and with similar severity scores. Then the major testing question is whether it is essential to examine the software under test for all the reported vulnerabilities with certain CVE numbers. It is possible to view the problem as an instance of general software testing problem and thus develop a specific adequacy criterion for covering vulnerabilities and examine the software under test (SUT) for the vulnerabilities. 

This paper extends our initial idea \cite{SAC2020} on vulnerability coverage and thus presents the novel concept of ``{\it vulnerability coverage},'' in an analogous way to conventional adequacy criterion in software testing. We apply evolutionary algorithms to generate test inputs with certain patterns and use Common Vulnerability Scoring System (CVSS) as a primary criterion to assess the vulnerability coverage (i.r., fitness) of the SUT. The paper makes the following key contributions:

\begin{enumerate}
    \item We present the idea of vulnerability coverage as a test adequacy criterion for inspecting the given software against certain types of vulnerabilities (Section \ref{sec:vulcov}).
    \item We perform evolutionary algorithms such as genetic algorithms (GA) and particle swarm optimisations (PSO) to generate vulnerability vector patterns (Section \ref{sec:algorithms}).%, as test inputs, and then filter out the set of vulnerabilities that match the generated vulnerability vector patterns to further examine the given software application.
    \item We compare the performance of both GA and PSO in generating such vulnerability vector patterns (Section \ref{sec:experiments}). According to our results, PSO managed to generate a more stable trend of secure vulnerability vector  patterns than that of GA in a single generation. %The results shows that the proposed adequacy criterion, i.e., vulnerability coverage, is effective for selectively choosing a certain class of vulnerabilities to further inspect the underlying applications.
\end{enumerate}
 
%In this paper, we will be using the words \textit{Vulnerability Vector/Pattern}, \textit{CVSS vector}, \textit{configurations} interchangeably.

%This paper is structured as follows: 
We provide relevant background on the Common Vulnerability Scoring System (CVSS) is presented in Section \ref{sec:CVSS}. %The background of the evolutionary algorithms used in this paper are presented in Section \ref{sec:evolutionary}. 
%We provide a definition for the idea of ``{\it vulnerability coverage}'' in Section \ref{sec:vulcov}.
%The adapted evolutionary algorithms are presented in Section \ref{sec:algorithms}. The experimental setup and results are reported in Section \ref{sec:experiments}. %The results of sensitivity analysis are reported in Section \ref{sec:sensitivity}. 
Section \ref{sec:vulcov} presents the idea of vulnerability coverage as an adequacy test criterion. Section \ref{sec:algorithms} presents the fitness function for the evolutionary algorithms. In Section \ref{sec:experiments}, we present experimental setup and results. Section \ref{sec:relatedwork} reviews the related work. Section \ref{sec:conclusion} concludes the paper. Throughout the paper, we will be using the words ``\textit{vulnerability pattern}'' and ``\textit{CVSS vector pattern}'' interchangeably.

\section{Vulnerability Scoring System}
\label{sec:CVSS}

The Common Vulnerability Scoring System (CVSS) is an open-standard industry framework, which helps cyber-security professionals to seek out information  regarding ranking mechanism for the severity of vulnerabilities. CVSS captures the principle characteristics of the vulnerability by assigning a Base score rating ranging from 0 to 10 which is representative of the ease of exploitation and the damaging effect of the concerned vulnerability where 10.0 is the most easily exploitable vulnerability. The numerical scores have a qualitative assessment (low, medium, high, and critical) to provide organizations with better understanding and assessment of vulnerabilities. Some vulnerability are also given temporal and environmental scores that may modify the base score. As a proof of concept in generating CVSS patterns, the GA and PSO optimization algorithms are applied to the Base metrics. The Base metric consists of three sub-main metrics where each metric group comprises of a set of vector fields and the associated values it takes:
\begin{itemize}
    \item \textbf{Exploitability Sub-Metric:} It addresses how the attack is captured. Table \ref{tab:EM-table} lists down the vector fields.% along with description and values each takes. 
    \item \textbf{Impact Sub-Metric:} It reflects the ``{\it characteristics}'' of the impacted components as shown in Table \ref{tab:EM-table}.  
    \item \textbf{Scope Sub-Metric:} It is a vector field acting as a separate metric which describes the change in the scope of the attack by determining whether other components are affected along with the original vulnerability. It accepts only two values: \textit{Unchanged (U)} and \textit{Changed (C)}.
\end{itemize}

\iffalse
\begin{figure}
\begin{center}
  \includegraphics[width=7cm]{cvss.png}
  \caption{The Base metric.}
  \label{fig:cvss}
  \vspace*{-0.2in}
  \end{center}
\end{figure}
\fi

\begin{table*}[h]
\centering
\begin{tabular}{|l|l|p{6cm}|c|l|}
\hline
%\rowcolor[HTML]{FFCCC9} 
\multicolumn{1}{|c|}{\textbf{Sub-Metric}} &
\multicolumn{1}{|c|}{\textbf{Fields}} &
  \multicolumn{1}{c|}{\textbf{Description}} &
  \multicolumn{1}{c|}{\textbf{Values}} &
  \multicolumn{1}{c|}{\textbf{Score}} \\ \hline
\rowcolor[HTML]{FFFFFF} 
Exploitability &  &  & Network(N)  & 0.85                                                                         \\ \cline{4-5} 
& &  & Adjacent(A) & 0.62                                                                         \\ \cline{4-5} 
&  &  & Local (L)   & 0.55                                                                         \\ \cline{4-5} 
& \multirow{-4}{*}{Attack Vector (AV)} &
  \multirow{-4}{*}{\begin{tabular}[c]{@{}l@{}}Reflects the proximity of the attacker to attack \\ the vulnerable component.\end{tabular}} &
  Physical (P) &
  0.2 \\ 
  \cline{2-5} 
&  &  & Low(L)      & 0.77                                                                         \\ \cline{4-5} 
& \multirow{-2}{*}{Attack Complexity (AC)} &
  \multirow{-2}{*}{\begin{tabular}[c]{@{}l@{}}Reflects the resources and conditions required to\\ conduct the exploit on the vulnerable component.\end{tabular}} &
  High(H) &
  0.44 \\ 
    \cline{2-5} 
&  &  & Low(L)      & \begin{tabular}[c]{@{}l@{}}0.62\\ (or 0.68 if Scope is Changed)\end{tabular} \\ \cline{4-5} 
&  &  & High(H)     & \begin{tabular}[c]{@{}l@{}}0.27\\ (or 0.5 if Scope is Changed)\end{tabular}  \\ \cline{4-5} 
& \multirow{-4}{*}{Privileges Required (PR)} &
  \multirow{-4}{*}{\begin{tabular}[c]{@{}l@{}}Represents the level of privileges required by an\\ attacker to successfully launch an exploit.\end{tabular}} &
  None(N) &
  0.85 \\ 
    \cline{2-5} 
&  &  & None(N)     & 0.85                                                                         \\ \cline{4-5} 
& \multirow{-2}{*}{User Interaction (UI)} &
  \multirow{-2}{*}{\begin{tabular}[c]{@{}l@{}}Reflects whether the participation of the user is\\ required for launching a successful attack.\end{tabular}} &
  Required (R) &
  0.62 \\ 
\hline
  Impact & Availability Impact (A) &
  Measures the severity of the attack on the availability of the impacted component. &
  Low(L) &
  0.00 \\ 
    \cline{2-5} 
& Integrity Impact (I) &
  Measures the severity of the attack on the integrity of the impacted component. &
  High(H) &
  0.22 \\ 
    \cline{2-5} 
& Confidentiality Impact (C) &
  Measures the severity of the attack on the confidentiality of the impacted component. &
  None(N) &
  0.56 \\ \hline
\end{tabular}
\caption{Sub-Metrics.}
\label{tab:EM-table}
\vspace{-0.15in}
\end{table*}

Base score formula is calculated as follows \cite{first}:

{
\scriptsize{
\begin{gather}
Impact\ Sub-Score(ISS) = 1 - [ (1 - C) * (1 - I) * (1 - A)]
\end{gather}
}
\vspace{-0.15in}
\scriptsize{
\begin{gather}
Impact(IM) =
  \begin{cases}
        6.42 * ISS & \text{if\textit{ Scope is Unchanged}}  \\
        7.52 * (ISS - 0.029) \\ 
        - 3.25 * (ISS - 0.02)^{15}  & \text{if \textit{Scope is Changed}}  \\
  \end{cases}
\end{gather}
}
\vspace{-0.15in}
\scriptsize{
\begin{gather}
Exploitability(EX) = 8.22 * AV * AC * PR * UI
\end{gather}
}
\vspace{-0.15in}
\scriptsize{
\begin{gather}
Base\ Score =
  \begin{cases}
        0 & \text{if $ Impact <= 0 $}
         \\
        Round(Min[(IM + EX), 10])                  & \text{if \textit{Scope is Unchanged}}  \\
        Round(Min[1.08 * (IM + EX), 10]) & \text{if \textit{Scope is changed}}  \\
  \end{cases}
  \end{gather}
}
\vspace{-0.15in}
}

\begin{figure}[!t]
  \includegraphics[width=8.5cm]{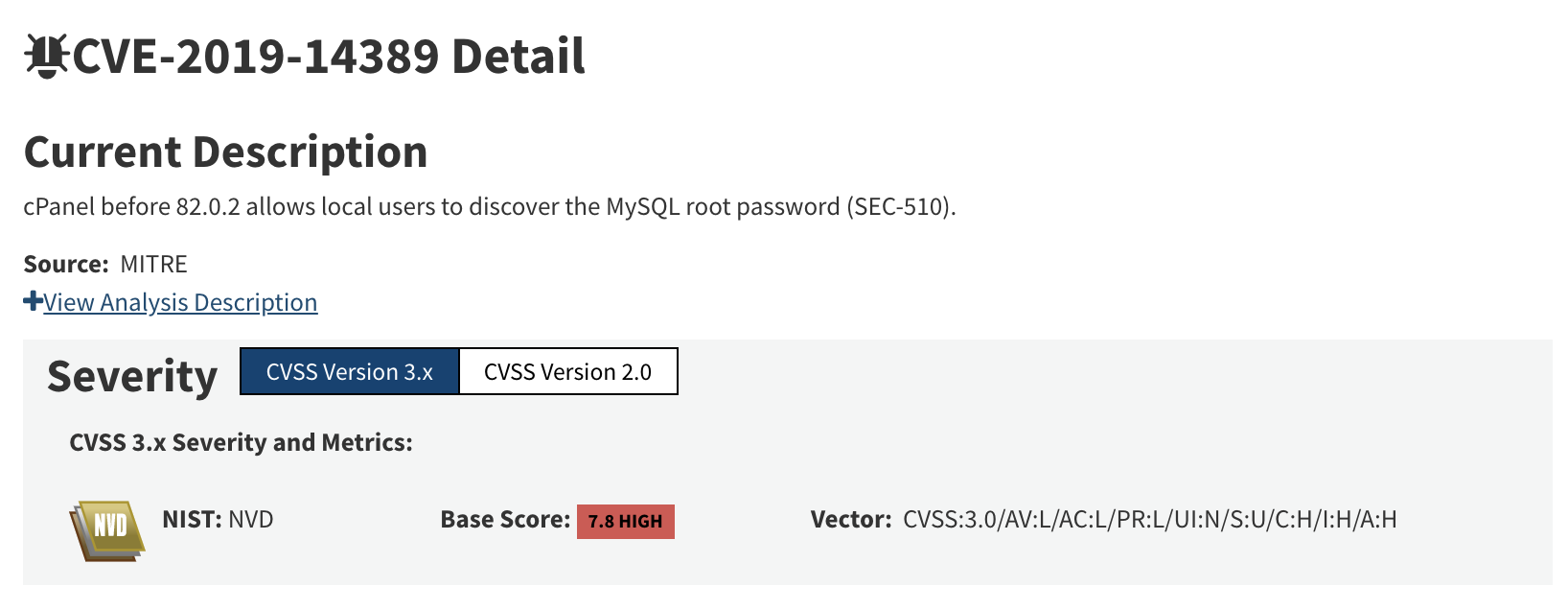}
  \caption{The description of CVE-2019-14389 for MySQL.}
  \label{fig:CVESample}
    \vspace*{-0.2in}
\end{figure}

%CVSS encapsulates all the Base sub\-metrics in a formula in order to quantify the vulnerability score. 
Every known vulnerability's severity can be represented as a vulnerability/CVSS vector pattern, which comprises all the aforementioned vector fields.
For instance, Figure \ref{fig:CVESample}  shows the CVSS score and the vulnerability pattern vector for CVE-2019-14389 \cite{NVD}. As shown in the figure, the score for this vulnerability is high and is quantified as $7.8$ out of 10. The generated representation of the vulnerability/CVSS vector is {\tt[AV:L/AC:L/PR:L/UI:N/S:U/C:H/I:H/A:H]}. Table \ref{tab:CVSSD} lists the vector pattern description for this vulnerability.

\begin{table}[h]
\centering
\scriptsize
\begin{tabular}{|c|p{6.5cm}|}
\hline
\multicolumn{1}{|c|}{\bf Parameter} & \multicolumn{1}{c|}{\bf Description} \\
\hline
AV: L & Denotes the vulnerability is exploited by the attacker through accessing the target system locally (L).        \\ 
\hline
AC: L & Represents that the vulnerability has a Low (L) complexity of being attacked.                         \\ 
\hline
PR: L & Shows that a Low (L) number of Privileges are required for successfully exploiting this vulnerability. \\ 
\hline
UI: N & Denotes that no (N) User Interaction  and involvement is required to launch a successful attack.     \\ 
\hline
S: U                   & Shows the Scope (S) of the attack is Unchanged (U).        \\ 
\hline
C: H                   & Total loss (High) of confidentiality.                             \\ 
\hline
I: H                   & Total loss (high) of integrity, or a complete loss of protection. \\ 
\hline
A: H  & Total loss (High) of availability, full access denial to resources in the impacted component.               \\ 
\hline
\end{tabular}
\caption{CVSS vector pattern description.}
\label{tab:CVSSD}
    \vspace*{-0.2in}
\end{table}

\section{Vulnerability Coverage As Adequacy Testing}
\label{sec:vulcov}

In an analogous way to the conventional definition given for ``code coverage'' in software testing, the vulnerability coverage is a measurement of how many known and reported vulnerabilities of the system under test (SUT) are inspected against. Similarly, the vulnerability coverage (VC) for a software system (S) can be measured as follows: 

\vspace*{-0.1in}

{
\scriptsize{
\begin{gather}
VC_{S} = \frac{(\#vulnerabilities\ inspected)_{S}}{(Total\ \#\ of\ vulnerabilities\ reported)_{S}} * 100 
\end{gather}
}
}
\vspace*{-0.10in}

Where $(\#vulnerabilities\ inspected)_{S}$ is the number of vulnerabilities inspected for the system $S$, and \\ $(Total\ \#\ of\ vulnerabilities\ reported)_{S}$ is the total number of vulnerabilities reported for system $S$. Without loss of generality, this paper uses {\it vulnerability patterns} provided by CVSS score to measure the adequacy testing of vulnerabilities give for a software system.

It is important to note that vulnerabilities of the same pattern might be found in CVE database directly. However, given the evolutionary search algorithms presented in this paper, it is possible to identify vulnerabilities of different patterns but similar CVSS score. Hence, the use of CVSS score, as a fitness function, enables us to identify various forms and patterns of vulnerabilities within the specific level of CVSS score. Therefore, given the desired level of CVSS score, the problem of adequacy testing for vulnerability testing will be exercising the vulnerabilities with different patterns but equal CVSS scores. In following sections, we adopt two evolutionary algorithms that enable us search the input space (i.e., vulnerability pattern) that achieve a certain level of CVSS score (i.e., the fitness value).

\section{Fitness Function: CVSS Score}
\label{sec:algorithms}

%Owing to the diverse nature of the parameters and the adaptability to optimization and search techniques such as evolutionary algorithms, we picked CVSS as a  prime choice in our paper to serve as a measure of fitness function to address this problem. Moreover, these types of greedy algorithm are a popular choice among other researchers  to address similar problems in test input generations in the context of random testing (e.g., \cite{DBLP:conf/kbse/AndrewsLM07}). 

%A system configuration is a collection of several parameters where each parameter have set of values/settings associated to it. Depending on what setting each parameter takes, a CVSS vector pattern is generated whose score reflects the vulnerabilities it embodies. The culmination of all the scores of the individual parameters amount to the  score of the configuration which tells us about the total number of possible vulnerabilities it contains along with its severity.
%Moreover, we can also account for the extent of diversity in the configurations by calculating how different their corresponding CVSS patterns are from each other. Hence, for this purpose we use Hamming distance to measure the differences between two patterns.

This section explains the genetic and optimization algorithms developed in which CVSS scores are used as fitness functions. For the ease of naming convention and understanding, we considered each CVSS pattern as a separate configuration in these algorithms.

%Each parameter in the configuration has a score vector associated to it which tells us about the vulnerability it contains along with its severity. The score is modeled into CVSS vector, which serves as a foundation to estimate the number of possible vulnerabilities of a certain configuration. Moreover, to determine the extent of diversity in the configuration generation, we also calculated the Hamming distance, which measures how different one pattern (i.e., configuration) is from another. This section presents the genetic and optimization algorithms developed based on the CVSS scores. 
\subsection{Genetic Algorithm (GA)}

Genetic Algorithms are based on the biological process of evolution. The idea is that over time, a pool of chromosomes will evolve to be even better (i.e., better fitness value) than the previous generation. A new generation (equal to the pool size) of chromosomes (i.e., configurations) is created with any iteration of the algorithm. This
is achieved by the processes of selection, crossover, and mutation\cite{GABook}. A fitness score metric is adopted as a measure to select the two
fittest chromosomes from the pool that are called parent chromosomes.
Then crossover takes place between the parents to produce
a new child chromosome, which will have the best traits from both
the parents followed by mutating of some of the characteristics of
the child to introduce new traits. This process is repeated until an
entirely new generation gets created.

\subsection{GA implementation for secure configuration pool}

We implemented the algorithm in Python. We first created a CVSS vector pool with the fitness score of $2.0$ (i.e., the best and more secure fitness score). We refer to vector as a ``string'' in our implementation. We set the number of iteration/generation as 50. The entire algorithm (Algorithm \ref{alg:GA}) is divided into five parts: 1) configuration generation , 2) fitness score, 3) breeder's Selection, 4) crossover, and  5) mutation. These parts are explained below in-depth. 

\subsubsection{Initial Configuration Generation} 
As shown on lines 1 -- 9 of Algorithm \ref{alg:GA}, we created a pool of 100 possible CVSS vector strings by randomly choosing corresponding permissible values from the `{\tt val}' list to produce the initial pool of vector strings.

\begin{algorithm}
		\caption{Genetic Alg. for generation of configurations.}
		\label{alg:GA}
		\begin{algorithmic}[1]
		    \State \Comment{Generating initial pool of configuration vectors.}
			\Procedure{Configuration}{}
			\State $val = {['H', 'L', 'N', 'A', 'P', 'U', 'C', 'N', 'R']}$
			\State $vector\_field= {['AV', 'AC', 'PR', 'UI', 'S', 'C', 'I', 'A']}$
			\For {each  $vf$ in $vector\_field$}
			\State $vf = random.choice[val]$ \Comment{'val' takes permissible set of values based on vf chosen.}
			\EndFor
			\State return $vector$ %\Comment{vector eg: AV:P/AC:H/PR:H/UI:R/S:U/C:N/I:N/A:H}
			\EndProcedure
			\State \Comment{ Assigning fitness score based on Best score.}
			\Procedure{Fitness}{$BestScore,vector$}
			\State $score = CVSS3(vector).score()$
			%\State $fit = 0$
			\If {{ $(score <= BestScore\ \&\ score <= 5.5)$}}
			\State $fit = score$
			\Else 
			\State $fit =100$
			\EndIf
			\State Return $fit$
			\EndProcedure
			\State \Comment{Breeder's Selection: Select best vector samples.}
			\Procedure{Selection}{$population, best\_sample, lucky\_few $}
			\State $nextGen = [\ ]$
			\State $sortedPop = Sorted(population)$ \Comment{descending order of fitness values = low cvss score to high}
			\For {$i$ in range(best\_sample)}
			\State $nextGen.append(sortedPop[i])$ \Comment{first 'i' \# of best configurations selected}
			\EndFor
			\For {$i$ in range(lucky\_few)}
			\State $nextGen.append(random.choice(sortedPop))$
			\EndFor
			\State Return $nextGen$
			\EndProcedure
			\State \Comment{Creating new vector from 2 parent vectors.}
			\Procedure{CreateChild}{$vector1,vector2$}
			\State $child\_vector =  "" $
		\For{ $i$ in range(len(vector1))}
			\If {{ $random.random<0.5$}}
			\State $child\_vector = child\_vector +vector1[i]$
			\Else 
			\State $child\_vector = child\_vector +vector2[i]$
			\EndIf
		 \EndFor
		     \State Return $child\_vector$
			\EndProcedure
			\State \Comment{Mutating: randomly changing a value of the vector.}
			\Procedure{Mutation}{$vector$}
			%\State $val = {['H', 'L', 'N', 'A', 'P', 'U', 'C', 'N', 'R']}$
			%\State $vectorField= {['AV', 'AC', 'PR', 'UI', 'S', 'C', 'I', 'A']}$
			\State $vf = random.choice(vector\_field)$
			\State $modify = random.choice(val)$
			\State $index = get\_position(vf)$
			\State $vector = vector[$:index$] + modify + vector[$index+1:$]$ \Comment{inserting 'modify' in the vector string}
			\State Return $vector$
			\EndProcedure
		\end{algorithmic}
\end{algorithm}

\subsubsection{Fitness Score} 
As shown on lines 11 -- 20 of Algorithm \ref{alg:GA}, the fitness score of the initial population of the vector strings is evaluated. We  imported the {\tt cvss} \cite{CVSS3} python library and thus utilized the base metric score method {\tt CVSS3}. The CVSS scores were considered as the fitness scores. The scores were valid if they were in the range of $[2.0, 5.5]$. Anything outside of that range was assigned the score as $100$. We chose score $5.5$ to be the upper limit since it is roughly the average score a configuration can take to be deemed reasonably secure.

\subsubsection{Breeder's Selection} 
As shown on lines 21 -- 32 of Algorithm \ref{alg:GA}, we then used Breeder's selection method. This method selects a combination of the best solutions generated by the algorithm (i.e., vectors with the low score). Furthermore, in order to avoid the problem of falling into local minima, the algorithm also picks some lucky few vectors with random vector scores. 

\subsubsection{Crossover} 
For crossover, As shown on lines 33 -- 44 of Algorithm \ref{alg:GA}, we randomly swapped the values of metrics among the two parent vectors. We used a random value generator to select which parent vector to use for crossover. If $value < 0.5$, parent 1 is chosen, otherwise parent 2 would be the choice.

\subsubsection{Mutation} 
As shown on lines 45 -- 53 of Algorithm \ref{alg:GA}, the algorithm performs mutation on the CVSS vector strings by random selection of  vector field whose value is also randomly selected from its permissible set of values. %Figure \ref{fig:finalpool} shows a view of the final generation of the vector strings along with their scores. As it is shown in the figure, the mutation procedure generates various combinations of the vector string whose fitness score is $2.0$. 

\iffalse{}
\begin{figure}
  \includegraphics[width=8.5cm]{scrn.png}
  \caption{The final pool of CVSS vector strings with their scores.}
  \label{fig:finalpool}
    \vspace*{-0.2in}
\end{figure} 
\fi

We ran the GA script 100 times. Each run of the algorithm produced different pool of CVSS vector strings with different number and combinations of vector of fitness score $2.0$. 

\subsection{Particle Swarm Optimization (PSO) Algorithm}

PSO is a widely used swarm-based optimization technique. It draws its inspiration from bee swarm, and bird flocking social behavior  of particles. PSO and GA, both being different forms of evolutionary computation techniques, share some similarities. Both techniques start off with a random set of initial population/solutions and keep updating generations until it reaches an optimum solution space with respect to the fitness function. In case of PSO, it is a swarm consisting of various particles, where each particle represents a solution. Unlike GA, PSO does not make use of crossover and mutation operators to update the particles. Instead, these techniques are directed towards the global optimum by their personal best position along with the swarm's best position in the search space. PSO is also easier to implement than GA and has comparatively fewer parameters to adjust \cite{PSOintro}.
\subsection{PSO implementation for secure pool configuration }

We compared the performance of GA in generating a set of best configurations with that of PSO. We implemented the PSO algorithm in Python 3.6. To make the comparison meaningful and fair, the number of iterations and population size (swarm size) were kept similar to GA, which are 50 and 100, respectively.

The PSO algorithm is described in Algorithm \ref{alg:PSO}. The algorithm takes two list as parameters: 1) {\tt pbest\_fitness} and 2) {\tt particle\_vel}. These lists maintain the initial {\tt pbest} fitness and velocity values associated to every particle in a swarm. We defined swarm as a collection (list) of 100 initial particles whose implementation (line 2) is similar to the procedure \texttt{configuration} in GA. The algorithm returns a pool of particles with varied scores in each iteration and also stores the count of particles with score = 2.0 in every iteration in order to check how many most secure particles (configurations) are generated by PSO. We also focus only on the scores, which belong in the range [2.0, 5.0] in every iteration. %The Particle Initialization (line 2) is similar to the configuration generation in GA. 

In a nutshell, the algorithm searches for the best fitness and velocity values for each particle until a threshold is reached (lines 7 -- 34). In every iteration (lines 9 -- 13), the algorithm picks the {\tt pbest\_fitness} (i.e., particle best) values as their CVSS scores {\tt cvss\_fit} with the assumption that the fitness would be better (i.e., lesser is better) than its current {\tt pbest} fitness value. After the {\tt For loop} ends, it then picks the global best ({\tt gbest})  value of the swarm by the the best {\tt pbest} value (lines 14 -- 16), in this case, the least value. 

The next step in the algorithm is to calculate the velocity (lines 17 - 31) where $particle\_vel(particle)$ fetches the velocity of the given particle. The lines 18 - 28 describe how the velocity is evaluated for every particle. The velocity metric measures the distance between the fitness score (pbest) and the best score.  The particle is updated whenever its current velocity value is greater than its previous one. 

The particles are updated using {\tt update\_particle} in a similar manner to the configuration mutation in GA. More specifically, any one out of the eight vector fields (i.e., AV, AC, etc.) is constructed whose value is chosen randomly from its corresponding set of permissible values. For example, if `AV'  is selected, then any value in the list of\{H, L, N, A\} can be randomly selected. 

The target global best value was set to $10.0$,  particle\_velocity in the range $[0, 8]$ where 0 and 8 are the minimum and maximum number of differences between two particles, respectively. Since each particle (CVSS vector) constitutes of only 8 vector fields (AV, AC, etc). The fitness\_range is set between the range $[2, 10]$ where $2.0$ is deemed as the best fitness score and 10.0 is the maximum CVSS score any particle can get which means highly unfit. 

\begin{algorithm}
      \caption{PSO for generation of configuration.}
		\label{alg:PSO}
		\begin{algorithmic}[1]
			\Procedure{PSO}{pbest\_fitness, particle\_vel}
			 \State swarm = [particle() for $i$ in range(swarm\_size)] \Comment{Initialize 100 particles}
			 \State iteration = 0
			 \State Threshold = 50
			 \State total\_count = [\ ] \Comment{To store count of particles with score = 2.0 in every iteration}
			 \State best\_score = 2.0 %\Comment{Target score to be achieved}
			 \While {iteration $<$ Threshold}
			 \State count = 0
			 \For {each particle} \Comment{Calc Fitness}
                    \If {$cvss\_fit(particle) <$ $pbest\_fitness(particle)$}
			           \State pbest\_fitness(particle) = cvss\_fit(particle)
			        \EndIf
			 \EndFor
			 \If {$pbest\_fitness(particle)$ $<$ $gbest(swarm)$}
			     \State gbest(swarm) = pbest(particle)
			 \EndIf
			 
			 \For {each particle} \Comment{Calc Velocity}
			          \If {$pbest\_fitness(particle)$$<$$best\_score$}
			              \State continue
			          \Else
    			            \State velocity=pbest(particle)-best\_score \Comment{Current Velocity}
    			            \If{$velocity == 0.0$}
			                    \State count+=1 \Comment{count of particles of score = 2.0}
			                 \EndIf
                         \If{$velocity<particle\_vel(particle)$}
                                \State $particle\_vel(particle) = velocity$
			               \Else
			                   \State particle = update\_particle(particle)
			               \EndIf
			         \EndIf
			 \EndFor
			 \State {$iteration += 1$}
			 \State {$total\_count.append(count)$}
			 \EndWhile
			 \State {return \space total\_count}
             \EndProcedure
	    \end{algorithmic}  
%	      \vspace*{-0.15in}
\end{algorithm}
%\vspace*{-0.15in}

\section{Experimentation and Results}
\label{sec:experiments}

%%%%%%%%%%%%%%%%%%%%%%%%%%%%%%%%%%%%%%%%%%%%%%%%%%%%%%%%%%%%%%%%%%%%%%%%%%%%%%%%
%\subsection{GA Performance}
We ran our Python scripts, developed for implementing the GA and PSO algorithms, 100 times on the CVSS population in order to evaluate the performance of the evolutionary algorithms in generating the most secure patterns. The performance was measured on the basis of three evaluation metrics:

\begin{enumerate}
    \item Number of instances of vulnerability patterns with the target score (e.g., score $ = (2.0, 3.0)$) in each run.  
    \item Mean hamming distance (diversity) of the CVSS vectors.
    \item Standard deviation of the scores calculated for the set of population produced.
\end{enumerate}

\subsubsection{Diversity of Vulnerability Patterns}

It is important to produce a diverse set of instances of vulnerability vector patterns to ensure the thoroughness of test inputs (i.e., vulnerability pattern) generation and thus avoid generating redundant test inputs where test input refers to an instance of vulnerability vector pattern produced by the algorithms. We collected the data for the three evaluation metrics for various range of target scores including \textit{S} $\in$ 2.0, \textit{S} $\in$ (2.0, 3.0], \textit{S} $\in$ (2.0, 4.0] \hspace{0.2cm} and \hspace{0.1cm} \textit{S} $\in$ (2.0, 5.0]. As a representative example, Figure \ref{fig:experimentGA} shows the plots for all the aforementioned three metrics for CVSS vector strings falling into \textit{S} $\in$ (2.0, 3.0] for both GA and PSO.

\iffalse
\begin{figure}%
    \centering
    \subfloat[Histogram Plot]{{\includegraphics[width=8.5cm]{cvss2.png} }}%
    \qquad
    \subfloat[Mean Hamming Distance ]{{\includegraphics[width=8.5cm]{cvss3.png} }}%
    \qquad
    \subfloat[Standard Deviation]{{\includegraphics[width=8.5cm]{cvss4.png} }}%
    \qquad
    \caption{Plot for metrics (a) Histogram, (b) Mean Hamming Distance, (c) Standard Deviation of targeted CVSS vector strings produced by GA where each plot in (a), (b) and (c) show results for target scores(S):
    \textit{S} $\in 2.0$ ,\textit{S} $\in (2.0, 3.0]$, \textit{S} $\in (2.0,4.0]$ \hspace{0.2cm} and \hspace{0.1cm} \textit{S} $\in (2.0,5.0]$.}
    \label{fig:experimentGA}%
      \vspace*{-0.2in}
\end{figure}

\begin{figure}%
    \centering
    \subfloat[Histogram Plot]{{\includegraphics[width=8.5cm]{pso1.png} }}%
    \qquad
    \subfloat[Mean Hamming Distance ]{{\includegraphics[width=8.5cm]{pso2.png} }}%
    \qquad
    \subfloat[Standard Deviation]{{\includegraphics[width=8.5cm]{pso3.png} }}%
    \qquad
    \caption{Plot for metrics (a) Histogram, (b) Mean Hamming Distance, (c) Standard Deviation of targeted CVSS vector strings produced by PSO where each plot in (a), (b) and (c) show results for target scores(S):
    \textit{S} $\in (2.0, 3.0]$, \textit{S} $\in (2.0,4.0]$ \hspace{0.2cm} and \hspace{0.1cm} \textit{S} $\in (2.0,5.0]$.}
    \label{fig:experimentPSO}%
      \vspace*{-0.2in}
\end{figure}
\fi

As expected, the number of instances of the generated vulnerability patterns for each run is smaller for the target score of $2.0$ (i.e., most secure) and it is higher when the target fitness score is in range $(2.0,5.0]$ (i.e., least secure). It implies that when a lower level of vulnerability is targeted (i.e., more secure with CVSS score $=2.0$), there are ``not'' too many alternatives for patterns. On the other hand, if the vulnerability levels and security is relaxed (i.e., CVSS score $<=5.0$ then over 60 alternatives could be produced for pattern matching. A combination of such varying level of CVSS scores might be beneficial to increase the search space when implementing a moving target defense platform. 

 The bar plots depicted in Figure \ref{fig:experimentGA}.(a) and \ref{fig:experimentGA}.(d) demonstrate  the number of occurrences of CVSS patterns (i.e., y-axis) against the number of runs (i.e., x-axis) when the target target CVSS scores is $(2.0,3.0)$ for GA and PSO, respectively. A glance at the charts indicates that GA is able to generate more instances of the CVSS patterns targeting the desired level of security (i.e., $(2.0, 3.0)$).

The scatter plots given in Figure \ref{fig:experimentGA}.(b)-(e) and Figure \ref{fig:experimentGA}.(c)-(f) denote distribution of the mean hamming distance and standard deviation against the runs, respectively. To ease comprehending the trend of the mean values, a regression line is fitted into the scatter plots to capture the overall trend. The Hamming distance addresses the ``{\it diversity}'' of the vulnerability vector patterns generated by the algorithms based on the count of corresponding unequal values of each vector fields among strings. The smoothing lines for mean values of Hamming distance demonstrate similar trends for each target value for the fitness score.

The mean values of the hamming distance (y-axis) in all the cases remain unchanged over the runs and  are mostly scattered between $3.0$ and $3.7$ for GA (i.e., a diversity of the vulnerability vector pattern generated) and between $4.5$ and $5.5$ for PSO. As demonstrated in scatter plots shown in Figure \ref{fig:experimentGA}.(b) and \ref{fig:experimentGA}.(c), the instances generated by the GA algorithm is less diverse compare to the instances generated by PSO. The mean of the hamming distances between the instances generated by GA and PSO are $3.45$ and $4.93$, respectively. This indicates that even though the PSO algorithm generates far fewer instances of CVSS patterns for the given fitness, it produces more diverse instances of patterns.

To illustrate the variations of such vulnerability patterns generated, plots \ref{fig:experimentGA}.(c) and \ref{fig:experimentGA}.(f) illustrate the trend of the values of the standard deviations for Hamming distance over the runs. There is a light reduction in standard deviations while running GA for all cases. The observed standard deviations for all cases is somewhere between $0.5$ and $1.5$. When combined together, the mean and standard deviations of the hamming distance can serve as an indication of the diversity of the vulnerability vector patterns produced by the algorithms and thus helps in measuring the thoroughness of test case generation and thus vulnerability selections in which the generation of redundant patterns (i.e., test inputs) is avoided.

\begin{figure*}%
    \centering
    \subfloat[Histogram Plot (GA)]{{\includegraphics[width=5.4cm]{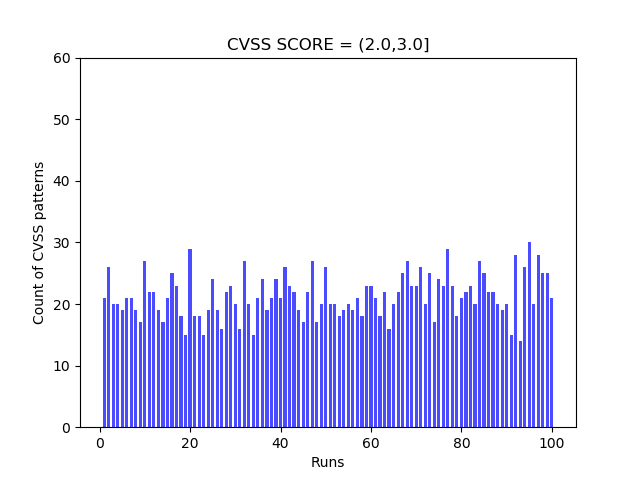} }}%
    \qquad
    \subfloat[Mean Hamming Distance (GA) ]{{\includegraphics[width=5.4cm]{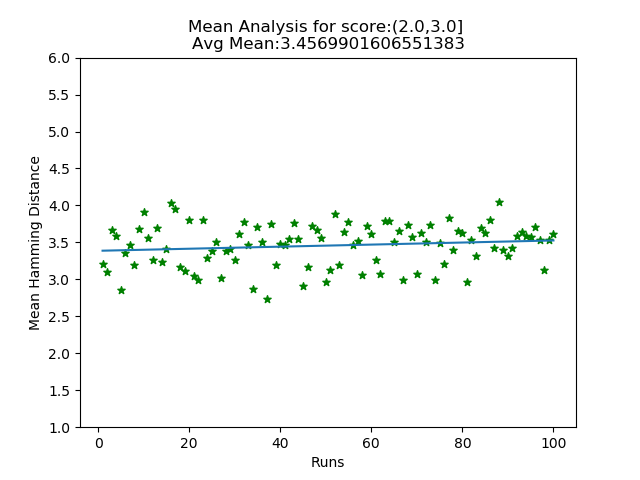} }}%
    \qquad
    \subfloat[Standard Deviation (GA) ]{{\includegraphics[width=5.4cm]{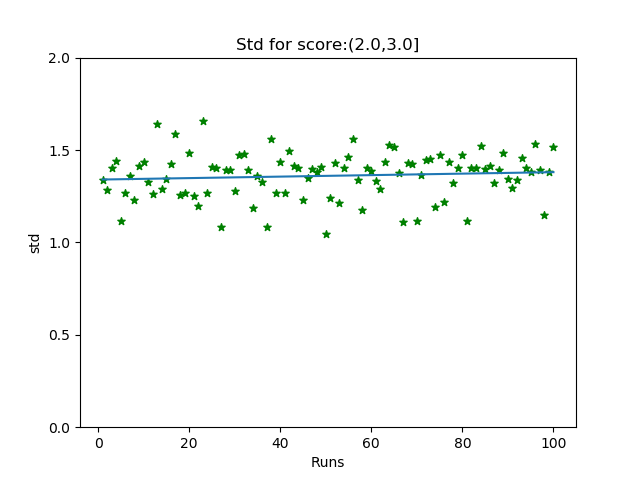} }}%
    \qquad
    \subfloat[Histogram Plot (PSO)]{{\includegraphics[width=5.4cm]{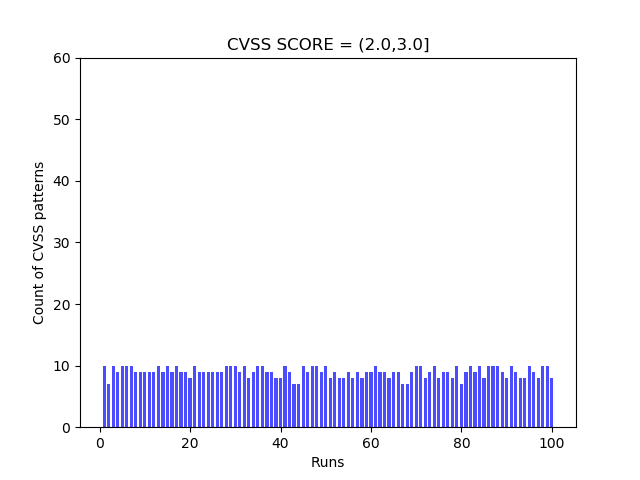} }}%
    \qquad
    \subfloat[Mean Hamming Distance (PSO) ]{{\includegraphics[width=5.4cm]{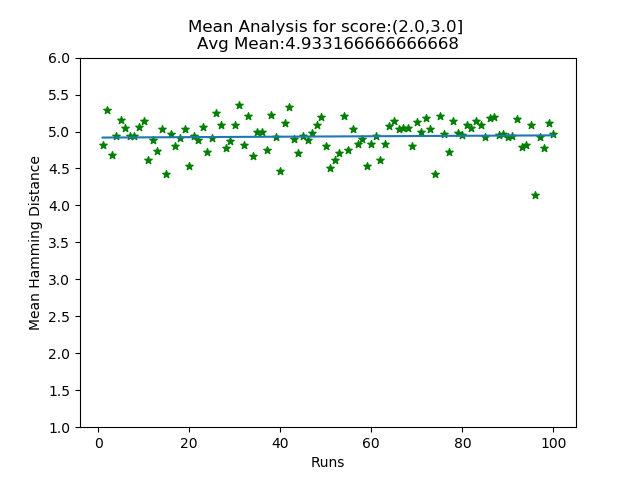} }}%
    \qquad
    \subfloat[Standard Deviation (PSO) ]{{\includegraphics[width=5.4cm]{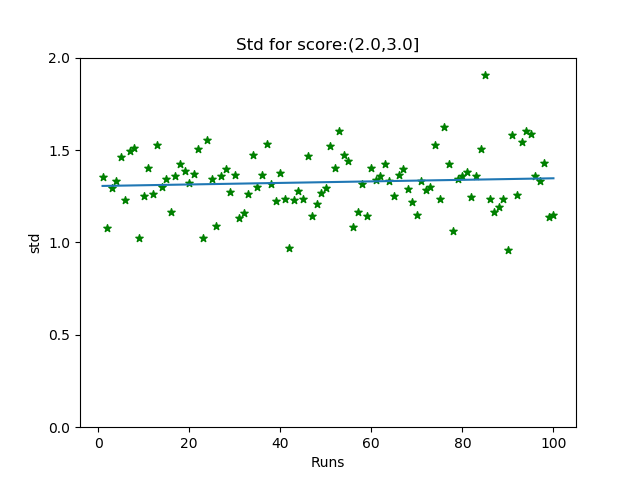} }}%
    \qquad
    \caption{Histograms, Mean Hamming Distances, Standard Deviations of CVSS vectors for \textit{S} $\in (2.0, 3.0]$.}
    \label{fig:experimentGA}%
      \vspace*{-0.2in}
\end{figure*}

\subsubsection{The Contributions of Each Permissible Value in each Vector Field}

It is also important to investigate whether certain settings of each vector field contributes to security configuration differently than its counterpart.  Table \ref{tab:GAvsPSO} shows the frequency (i.e., in terms of percentage) of each value permissible for each vector field, as listed in the base metrics. As reported in Table \ref{tab:GAvsPSO}:

\begin{itemize}
    \item[--]{\bf AV}: The most contributing value is P (i.e., Physical) for  GA (ranging from $45.43$ to $59.22\%$). The PSO algorithm highlights two values of P and L as the most contributing to the security level of the patterns. This observation indicates that if the severity of the vulnerability needs to be reduced, no other values or means of attacks (i.e., Network (N), Adjacent (A), and somewhat Local (L) is allowed for exploiting the vulnerability. 
    
    \item[--]{\bf AC}: There is a mixed situation for attack complexity and there is no clear winner between Low (L) and High (H) complexity level to launch the exploitation. 
    
    \item[--]{\bf S}: The dominant setting for this variable is C, except the case for GA when the target score is $2.0$.    
    \item[--]{\bf UI}: There is a mixed situation for the level of user involvement for exposing the vulnerability. 
    
    \item[--]{\bf C}: There is also a mixed situation for confidentiality settings among GA and PSO algorithms.   
    
    \item[--]{\bf I}: A similar mixed situation for this case. However, it is also observed that in most cases a None (N) risk to integrity is needed to reduce the impact of exploiting the vulnerability. 
    
    \item[--]{\bf A}: Furthermore, there is a a mixed situation  for availability where there is no clear dominant setting value.
    
     \item[--]{\bf PR}: The two dominant setting values for the level privileges are L and H.
\end{itemize}

\begin{figure}%
    \centering
    \subfloat[GA]{{\includegraphics[width=8cm]{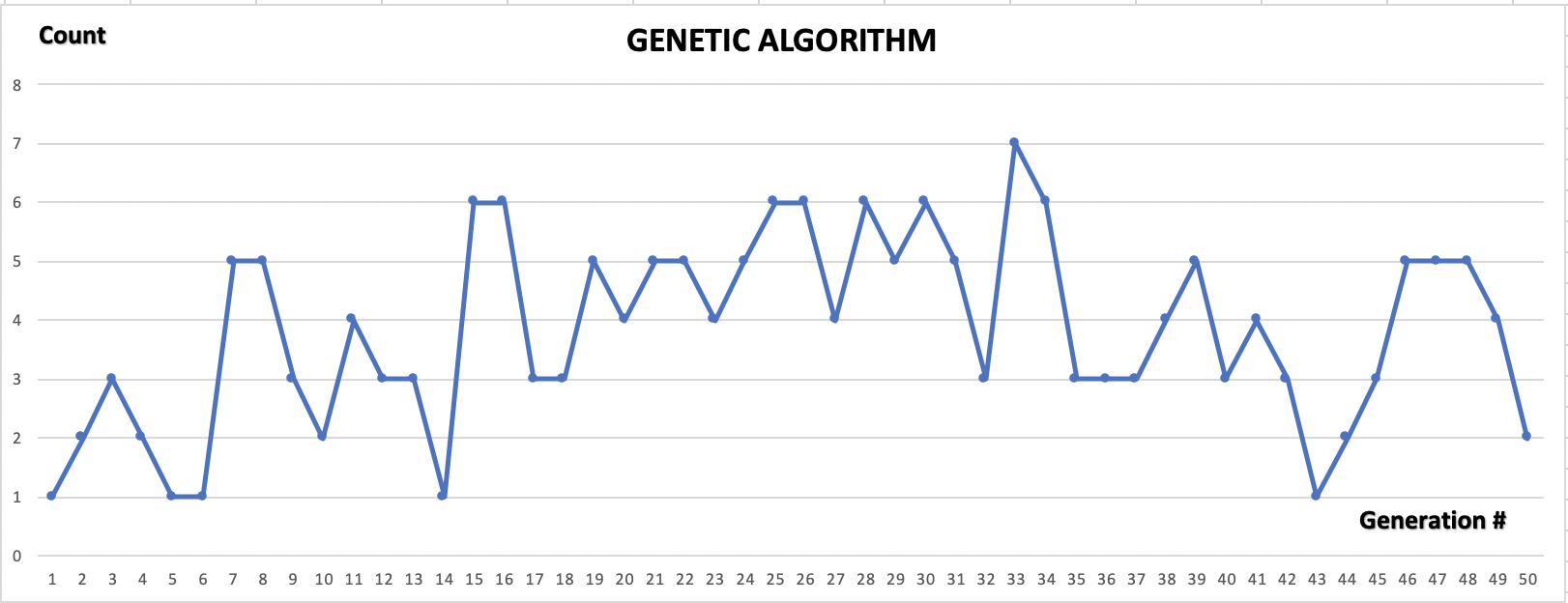} }}%
    \qquad
    \subfloat[PSO]{{\includegraphics[width=8cm]{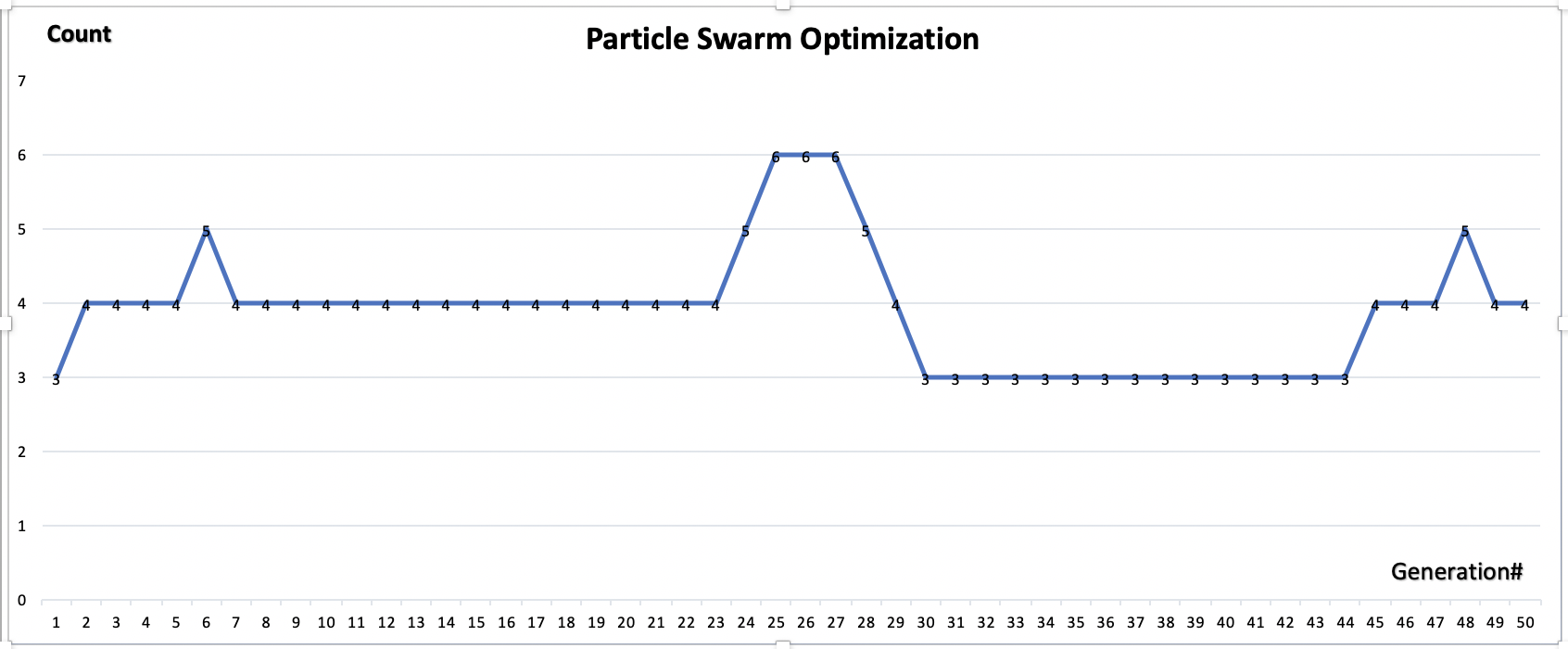} }}%
    \qquad
    \caption{\# CVSS patterns with score 2.0 over 50 generations.}
    \label{Fig:singlerunGAPSO}
      \vspace*{-0.25in}
\end{figure}

We also executed the GA and PSO scripts with single run to sense their performances. Figure \ref{Fig:singlerunGAPSO} illustrates the results of one run with 50  generations/iterations where y-axis is the number of vulnerability pattern produced whose score = 2.0; whereas, the x-axis is the generation index (i.e., 50 generations). As it is observed, GA could manage to generate two vulnerability vector patterns with score 2.0 on its 50-$th$ iteration along with highly unstable trend; whereas, the PSO algorithm demonstrated a more stable trend with four pattern generated with score 2.0.  

\begin{table*}[!t]
\centering
\begin{tabular}{|l|l|l|l|l|l|l|l|l|l|}
\hline
\textbf{Vector} &
  \textbf{Values} &
  \multicolumn{2}{c|}{\textbf{[2.0]}} &
  \multicolumn{2}{c|}{\textbf{(2.0, 3.0]}} &
  \multicolumn{2}{c|}{\textbf{(2.0, 4.0]}} &
  \multicolumn{2}{c|}{\textbf{(2.0, 5.0]}} \\ 
    \cline{3-10} 
 \textbf{Field} &
  \textbf{} &
  \textbf{GA} &
  \textbf{PSO} &
  \textbf{GA} &
  \textbf{PSO} &
  \textbf{GA} &
  \textbf{PSO} &
  \textbf{GA} &
  \textbf{PSO} \\ 
\hline
\hline
                              & P                        &{\bf 59.22}  &{\bf 29.29}  & {\bf 49.31}  &23.74  &{\bf 47.85}  &23.19  &{\bf 45.43}  &22.25  \\ \cline{2-10} 
                              & L                        &14.02  &22.22  &21.40  &{\bf 26.75}  &21.90  &{\bf 26.98}  &22.32  &{\bf 26.95}  \\ \cline{2-10} 
                           & A                        &13.76  &26.26  &18.67  &26.15 &18.41  &25.25  &19.11  &25.33  \\ \cline{2-10} 
\multirow{-5}{*}{\textbf{AV}} & N                       &12.98  &22.22  &10.60 &23.34  &11.83  &24.56  &13.13 &25.45  \\ \hline \hline
                              & L  &32.98  &44.44  &48.63  &{\bf 50.90}  &{\bf 50.83}  &{\bf 50.17}  &{\bf 51.85}  &{\bf 51.56}  \\ \cline{2-10} 
\multirow{-2}{*}{\textbf{AC}}          &H  &{\bf 67.01}  &{\bf 55.55}  &{\bf 51.36}  &49.09  &49.16  &49.82  &48.14  &48.4  \\ \hline \hline
                             &U                          &{\bf 60.25}  &48.48  &47.87  &48.39  &47.15  &48.96  &46.50  &48.56  \\ \cline{2-10} 
\multirow{-2}{*}{\textbf{S}}           & C                         &39.74  &{\bf 51.51}  &{\bf 52.12}  &{\bf 51.60}  &{\bf 52.84}  &{\bf 51.03}  &{\bf 53.49}  &{\bf 51.43}  \\ \hline \hline
                             &N  &33.24  &43.43  &48.51  &{\bf 50.0}  &49.30  &{\bf 50.40}  &49.79  &{\bf 50.70}  \\ \cline{2-10} 
\multirow{-2}{*}{\textbf{UI}}          &R  &{\bf 66.75}  &{\bf 56.56}  &{\bf 51.48}  &{\bf 50.0}  &{\bf 50.69}  &49.59  &{\bf 50.20}  &49.29  \\ \hline \hline
                              &N                          &{\bf 62.85}  &25.25  &{\bf 61.84}  &31.46  &{\bf 54.77}  &29.16  &{\bf 53.36}  &28.88  \\ \cline{2-10} 
                              &L                          &37.14  &{\bf 37.37}  &38.15  &{\bf 35.67}  &45.19  &{\bf 36.73}  &45.82  &34.91  \\ \cline{2-10} 
\multirow{-3}{*}{\textbf{C}}           &H                         &0.0  &{\bf 37.37}  &0.0  &32.86  &0.02  &34.09  &0.80  &{\bf 36.20}  \\ \hline \hline
                              &N  &{\bf 67.27}  &{\bf 39.39}  &{\bf 64.09}  &30.46  &{\bf 55.98}  &27.21  &{\bf 54.54}  &25.84  \\ \cline{2-10} 
                             &L  &32.72  &34.34  &35.90  &{\bf 38.47}  &43.82  &{\bf 37.88}  &44.37  &35.72  \\ \cline{2-10} 
\multirow{-3}{*}{\textbf{I}}           &H  &0.0  &26.26  &0.0  &31.06  &0.18  &34.90  &1.07  &{\bf 38.42}  \\ \hline \hline
                              &N                          &{\bf 69.87}  &26.26  &{\bf 66.70}  &28.05  &{\bf 58.96}  &27.95  &{\bf 57.95}  &27.17  \\ \cline{2-10} 
                              &L                          &30.12  &{\bf 47.47}  &33.29  &{\bf 37.17}  &40.98  &34.95  &41.55  &34.01  \\ \cline{2-10} 
\multirow{-3}{*}{\textbf{A}}           &H                          &0.0  &26.26  &0.0  &34.76  &0.05  &{\bf 37.08}  &0.49  &{\bf 38.81}  \\ \hline \hline
                              &N  &40.25  &26.26  &29.75 &28.05  &30.81  &27.95  &30.58  &27.17  \\ \cline{2-10} 
                              &L  &0.0  &{\bf 47.47}  &29.83  &{\bf 37.17}  &31.13  &34.95  &31.70  &34.01  \\ \cline{2-10} 
\multirow{-3}{*}{\textbf{PR}}          &H  &{\bf 59.74}  &26.26  &{\bf 40.40}  &34.76  &{\bf 38.05}  &{\bf 37.08}  &{\bf 37.71}  &{\bf 38.81}  \\ \hline  
\end{tabular}
\caption{\% of contribution of each permissible value in all the score ranges across 100 runs of GA and PSO.}
\label{tab:GAvsPSO}
\vspace{-0.2in}
\end{table*}

\section{Related Work}
\label{sec:relatedwork}

Crouse and Fulp \cite{6111663} used genetic algorithms to deploy a Moving Target Defense (MTD) platform and make computer systems more secure through temporal/spatial diversity in configuration parameters that govern how a system operates. Later on, they developed an MTD by simulating 256 virtual machines of similarly purposed computers where each computer was initially configured with an extremely vulnerable configuration making them prone to all sorts of attacks. %Every configuration in a simulated computer consisted of 80 parameters, half of which belonged to security. The chromosome pool size, crossover and mutation rates were initially kept 10, 0.08 and 0.02 respectively. %Through this experiment, Crouse and Fulp reported that both Average temporal and spatial diversity first increase and then reduce and remain constant with increase in number of iterations.

Post and Sinz \cite{Post:2008:CLV:1642931.1642971} bridged the gap between configuration information and verification process by introducing a new technique named \textit{Configuration Lifting}. The technique converts all the variants over which a software is verified into a meta-program thereby making the application of configuration-aware verification techniques like static analysis, and model checking more efficient.% This technique is primarily used for building configurations and compilation such that all the configuration steps are performed at run time. The authors also checked the feasibility of this technique by applying it on the configuration dependent hazards of Linux Kernel, which houses several thousands of configurable features and successfully found out two novel bugs in the system.
Dai et al.\ \cite{5438043} introduced the concept of \textit{configuration fuzzing} in order to check the vulnerabilities that appear only at certain conditions by randomly modifying the configuration of the running application at specific execution points. During the deployment phase, this technique ceaselessly fuzzes the configuration and looks for a vulnerability that rises due to the violation of of security invariants. %This process takes place in a duplicate copy of the application to avoid the original running application from hampering.

%      \vspace*{-0.1in}
\section{Conclusion}
\label{sec:conclusion}
We introduced the novel idea of ``vulnerability coverage,'' a methodology to examine software under test against certain classes of vulnerabilities as reported by National Vulnerability Database (NVD) adequately. The introduced idea makes use of an open industry standard tool called Common Vulnerability Scoring System (CVSS) as a metric to measure fitness in order to generate  a pool of vulnerability vector patterns that attains a secure level of CVSS score. For adequacy testing of the underlying software, the software under test is then inspected against all  those  filtered representative sets of vulnerabilities with similar vulnerability vector pattern that were selected from the generated pool. The paper compared two evolutionary-based algorithms namely Genetic and Participle Swarm Optimization algorithms on the basis of their performance in generating a pool of vulnerability patterns and the results indicated a similar performance achieved by both algorithms.

The concept of adequacy criterion is a new approach and hence has a larger scope of improvement. An adequacy criterion based on vulnerability coverage is a novel technique in the best of our knowledge. This approach can be further improved by taking into consideration several other metrics including temporal and environmental ones present in  CVSS and National Vulnerability Database (NVD). We also built our experiments based on the range of $2.0$ and $5.5$. Additional experimentation would be needed to further study the effect of such range. Moreover, the concept needs tool support and further empirical studies which can aid in thorough and systematic searching for vulnerabilities reported in the NVD database based on the matching property for the goal of security testing and then investigate the effectiveness of such adequacy criterion.

\iffalse
This paper introduced the concept of ``vulnerability coverage'' as an adequacy criterion for security and vulnerability testing of software applications. the deriving idea is to utilize Common Vulnerability Scoring System (CVSS) as a fitness metric and identify a set of vulnerability vector patterns that achieves a certain level of CVSS score. The generated set can be used for adequacy testing of underlying software in which all or representative sets of vulnerabilities with similar vulnerability vector pattern will be selected for further inspecting the software under test against for. The paper compared two evolutionary-based algorithms namely Genetic and Participle Swarm Optimization algorithms and the results indicated a similar results obtained by both algorithms. 

The novel idea of adequacy criterion as introduced in this paper needs further attention. To our best knowledge, an adequacy criterion based on vulnerability coverage does not exist. There are several other features that need to investigated including other metrics incorporated into CVSS and National Vulnerability Database (NVD) including temporal and environmental metrics. Furthermore, the idea needs tool support and further empirical studies to thoroughly search the NVD database for reported vulnerabilities with exact pattern matching property for security testing purposes and investigate the effectiveness of such adequacy criterion. 
\fi

%\section*{Acknowledgment}
%\twocolumn
\section*{Acknowledgment}
This work is supported in part by funding from National Science Foundation under grants no: 1516636 and 1821560.

\iffalse

\fi

\bibliography{References} 
\bibliographystyle{plain}
\end{document}